\documentclass{article}

\usepackage{arxiv}

\usepackage[utf8]{inputenc} 
\usepackage[T1]{fontenc}    
\usepackage{hyperref}       
\usepackage{url}            
\usepackage{booktabs}       
\usepackage{amsfonts}       
\usepackage{nicefrac}       
\usepackage{microtype}      
\usepackage{lipsum}		
\usepackage{graphicx}
\usepackage{natbib}
\usepackage{doi}

\title{Fine-tune the Entire RAG Architecture (including DPR retriever) for Question-Answering}

\author{ \href{}{{}\hspace{1mm}Shamane Siriwardhana} \\
	Auckland Bioengineering Instutute\\
	The University of Auckland\\
	Auckland, New Zealand\\
	\texttt{shamane@ahlab.org} \\
	\And
	\href{}{{}\hspace{1mm}Rivindu Weerasekera} \\
	The University of Auckland\\
	Auckland, New Zealand\\
	\texttt{rivindu@ahlab.org} \\	

	\And
	\href{}{{}\hspace{1mm}Elliott Wen} \\
	Auckland Bioengineering Instutute\\
	The University of Auckland\\
	Auckland, New Zealand\\
	\texttt{elliott@ahlab.org} \\
	
	\And
	\href{}{{}\hspace{1mm}Suranga Nanayakkara} \\
	Auckland Bioengineering Instutute\\
	The University of Auckland\\
	Auckland, New Zealand\\
	\texttt{suranga@ahlab.org} \\	

}

\hypersetup{
pdftitle={A template for the arxiv style},
pdfsubject={q-bio.NC, q-bio.QM},
pdfauthor={David S.~Hippocampus, Elias D.~Striatum},
pdfkeywords={First keyword, Second keyword, More},
}

\begin{document}
\maketitle
\begin{abstract}
In this paper, we illustrate how to fine-tune the entire Retrieval Augment Generation (RAG) architecture  ~\citep{lewis2020retrieval} in an end-to-end manner. We highlighted main engineering challenges that needed to be addressed to achieve this objective. We also compare how end-to-end RAG architecture outperforms the original RAG architecture for the task of question answering. We have open-sourced our implementation in the \textbf{HuggingFace Transformers library}~\citep{wolf2019huggingface}. 

\url{https://github.com/huggingface/transformers/tree/master/examples/research_projects/rag-end2end-retriever}.
\end{abstract}


\section{Introduction}
In September 2020, Facebook open-sourced a new NLP model called Retrieval Augmented Generation (RAG) on the Hugging Face Transformer library. RAG is capable to use a set of support documents from an external knowledge base as a latent variable to generate the final output. The RAG model consists of an Input Encoder, a Neural Retriever, and an Output Generator. All three components are initialized with pre-trained transformers. However, the original Hugging Face implementation only allowed fine-tuning the Input Encoder and the Output Generator in an end-to-end manner, while the Neural Retriever needs to be trained seperately. To the best of our knowledge, an end-to-end RAG implementation that trains all three components does not exist. In this paper, we introduce a novel approach to extending the RAG implementation which fulfills the task of training all three components end-to-end. Although this appears straightforward, there are many engineering challenges that need to be addressed.

As the name suggests, RAG adds an intermediate information retrieval step before the final generation process. It combines the ideas of neural information retrieval with seq2seq generation in an end-to-end manner. During the training phase, RAG takes an input and encodes it to a high-dimensional vector. This vector then gets used as a query to select the most relevant passages from an external database using Maximum Inner Product Search (MIPS)~\citep{JDH17,guo2020accelerating}. Afterward, The input and the selected set of documents get fed into a generator that produces the final answer. It is important to emphasize that components of RAG are being initialized with pre-trained BERT and BART transformers. Finally, The probability of selecting documents given an input context ($p(z|x)$) is used to propagate gradients back to the question encoder. During the process~ \footnote{\href{https://shamanesiri.medium.com/how-to-finetune-the-entire-rag-architecture-including-dpr-retriever-4b4385322552}{Please refer to our medium tutorial by clicking this link}}. of making the RAG end-to-end trainable, we mainly work on the Retriever part, which uses a neural retriever model called Dense Passage Retrieval (DPR)~\citep{karpukhin2020dense}.

Our main contributions can be summarized as follows:

\begin{itemize}
  \item Highlighting the importance of making entire RAG architecture end-to-end trainable.
  \item Exploration of engineering challenges regarding the extension of original RAG.
  \item Comparing the original RAG with end-to-end RAG on question answering to highlight the effectiveness of making entire RAG end-to-end trainable.
  \item Open source implementation
\end{itemize}

\section{Rag Retriever}

In our extension, we mainly work with the RAG retriever. The retriever basically consists of a pre-trained DPR model~\citep{karpukhin2020dense}. The DPR model consists of two BERT models. One model encodes questions and the other model encodes the documents using the CLS token outputs. The DPR model used in RAG has already been trained with passages and questions extracted from open domain Wikipedia-based datasets. RAG has a neural retriever and a reader combined in an end-to-end manner. In practice, however, we freeze the passage encoder and only train the question encoder. In our approach, the passage encoder is trainable as well.
In RAG, prior to fine-tuning, a frozen Passage Encoder is used to encode your external Knowledge Base (KB) and save it to disk. Then we do the dataset indexing. Imagine there are millions of documents that you need to search and retrieve during the training time. Usual vector similarity can be too slow. So we use some approximation methods by clustering the vector space into sub-regions. There are several libraries for this, such as ScANN~\citep{guo2020accelerating} and FAISS~\citep{JDH17}. In the RAG implementation Huggingface uses FAISS to make the retrieval phase faster. After this step, we start the training process with the indexed dataset where we only update the model parameters of the Question Encoder and Generator networks. We do not use the Passage Encoder or change the encoded embeddings for documents during the training.

\subsection{Why is it important to fine-tune the entire retriever?}

RAG authors illustrated that it is acceptable not to fine-tune the Passage Encoder for tasks like question answering and fact-checking. But the authors have conducted their experiments mainly on open domain Wikipedia-like datasets~\citep{kwiatkowski2019natural}. Since DPR is also initially trained on Wiki-data, this really makes sense. Our exploration is mainly motivated by the following factors.

\begin{itemize}
  \item Does training of the entire RAG-retriever help with domain adaptation?
  \item The authors of similar retrieval augmented models like REALM~\citep{guu2020realm,sachan2021end} mention it would be advantageous to train the entire retriever.
\end{itemize}

\section{Why is it hard to make the entire RAG end-to-end trainable?}

Theoretically, there are no problems in propagating gradients to both the passage encoder and question encoder BERT model. As described in the RAG loss function~\citep{lewis2020retrieval}, we can compute the document scores for latent documents (p(z|x)) by computing question encoder embeddings and document encoder embeddings during the training time. This has the following two steps,

\begin{itemize}
  \item Combined a pre-trained Passage Encoder model to the RAG model prior to the training.
  \item During the training, first, retrieve relevant passages given a question by using the indexed dataset.
  \item Calculate the document scores by re-computing CLS embeddings for retrieved documents using the initialized passage BERT model and do the same thing for input using the question encoder.
\end{itemize}

Although the propagation of the gradients to the passage encoder is straightforward, we need to make sure the updated passage encoder gets used in the overall training architecture. So we have to update the indexed dataset during the training process.

\subsection{Why updating the indexed KB during the training is time-consuming and computationally expensive?
}

The main two engineering challenges can be illustrated as follows.

\begin{itemize}
  \item During training, we first need to send each passage in the external KB through the updated passage encoder to compute the CLS token. Let’s call this step re-encoding the KB.
  \item Secondly, with the updated embeddings for each passage, we need to re-index the external KB with FAISS. Let’s call this part re-indexing of the KB.
\end{itemize}

So, similar to the REALM training process, we can execute external KB re-encoding and re-indexing commands every N-training steps. The authors in REALM have mentioned that if we improve the frequency of KB updates, we can get better results (check section 3.3 in REALM paper~\citep{kwiatkowski2019natural}). Since the re-encoding and re-indexing processes can take large amounts of execution time, we should not practically hold the training process for hours until we finish the above two processes (read about the effectiveness of stale gradients in REALM paper in section 3.3).

\section{Making the re-encoding and re-indexing practical during the training loop}

We achieve this task by using HuggingFace Datasets, Pytorch-Lightning~\cite{falcon2019pytorch}, and Python Multiprocessing libraries. First, we need to make sure the re-encoding step is fast and does not make the entire training loop wait until it finishes the computation of embeddings for every passage in the KB. To achieve this we executed the re-encoding and re-indexing processes parallel to the main training loop. 

To have a detailed explanation about the implementation details and a code walk-through, \href{https://shamanesiri.medium.com/how-to-finetune-the-entire-rag-architecture-including-dpr-retriever-4b4385322552}{\textbf{please refer our medium article}}\footnote{\url{https://shamanesiri.medium.com/how-to-finetune-the-entire-rag-architecture-including-dpr-retriever-4b4385322552}}.

\section{Results}

We conducted a simple experiment to investigate the effectiveness of this end2end training extension using the SQuAD dataset. First we Created a knowledge-base using all the context passages in the SQuAD dataset with their respective titles. It is important to note that we chunk each context in to maximum of 100 words, which creates around 20 000 passages. 
Then we used the standard training and validation splits to train and evaluate our model.  As illustrated in Table 1, our results suggest that finetuning the RAG with the entire-DPR can outperform the original RAG by 12\%.

\begin{table}[h]
\begin{tabular}{|l|l|}
\hline
Model Name   & Exact Match \\ \hline
RAG-Original & 28.12       \\ \hline
RAG - (Ours) & 40.02       \\ \hline
\end{tabular}
\caption{Comparison of two RAG models on SQUAD training data, when using contexts as the knowledge base. }
\label{tab:my-table}
\end{table}

\bibliographystyle{unsrtnat}
\bibliography{references}  






\end{document}